\documentclass[preprints,article,accept,moreauthors,pdftex,10pt,a4paper]{mdpi}
\usepackage{graphicx}
\usepackage{placeins}

\usepackage{amsfonts}
\usepackage{amssymb}
\usepackage{dsfont}
\usepackage{relsize}
\usepackage{multirow}
\usepackage{hyperref}
\usepackage[utf8]{inputenc}
\usepackage[T1]{fontenc}
\usepackage{newtxtext,newtxmath}
\firstpage{1} 
\pubvolume{xx}
\issuenum{1}
\articlenumber{5}
\pubyear{2018}
\copyrightyear{2018}
\history{}
\Title{Exploiting Convexification for Bayesian Optimal Sensor Placement by Maximization of Mutual Information}

\Author{Pinaky Bhattacharyya $^{1,\dagger,\ddagger}$\orcidA{} and James Beck $^{2,}$*}

\AuthorNames{Pinaky Bhattacharyya and James Beck}

\address{%
$^{1}$ \quad Praedicat, Inc.; formerly Graduate Student, California Institute of Technology; pinaky\_b@alumni.caltech.edu\\
$^{2}$ \quad California Institute of Technology; jimbeck@caltech.edu}

\corres{Correspondence: pinaky\_b@alumni.caltech.edu; Tel.: +1-626-399-8070}

\abstract{Bayesian optimal sensor placement, in its full generality, seeks to maximize the mutual information between uncertain model parameters and the predicted data to be collected from the sensors for the purpose of performing Bayesian inference. Equivalently, the expected information entropy of the posterior of the model parameters is minimized over all possible sensor configurations for a given sensor budget. In the context of structural dynamical systems, this minimization is computationally expensive because of the large number of possible sensor configurations. Here, a very efficient convex relaxation scheme is presented to determine informative and possibly-optimal solutions to the problem, thereby bypassing the necessity for an exhaustive, and often infeasible, combinatorial search. The key idea is to relax the binary sensor location vector so that its components corresponding to all possible sensor locations lie in the unit interval. Then, the optimization over this vector is a convex problem that can be efficiently solved. This method always yields a unique solution for the relaxed problem, which is often binary and therefore the optimal solution to the original problem. When not binary, the relaxed solution is often suggestive of what the optimal solution for the original problem is. An illustrative example using a fifty-story shear building model subject to sinusoidal ground motion is presented, including a case where there are over 47 trillion possible sensor configurations. The solutions and computational effort are compared to greedy and heuristic methods.}


\begin{document}
%%%%%%%%%%%%%%%%%%%%%%%%%%%%%%%%%%%%%%%%%%
%% Only for the journal Gels: Please place the Experimental Section after the Conclusions

%%%%%%%%%%%%%%%%%%%%%%%%%%%%%%%%%%%%%%%%%%
%\setcounter{section}{-1} %% Remove this when starting to work on the template.
\newcommand{\data}{\mathcal{D}}
\newcommand{\model}{\mathcal{M}}
\newcommand{\normal}{\mathcal{N}}
\newcommand{\gammainv}{\Gamma^{-1}}
\newcommand{\datad}{\data(\delta)}
\newcommand{\likelihood}[1]{p(#1|\theta,\model)}
\newcommand{\prior}{p(\theta|\model)}
\newcommand{\posterior}[1]{p(\theta|#1 \model)}
\newcommand{\evidence}[1]{p(#1|\model)}
\newcommand{\newdata}{Y'_{N'}(\delta')}
\newcommand{\olddata}{Y_N(\delta)}
\newcommand{\datavarn}{y_{1:N}}
\newcommand{\datavarnhat}{\hat{y}_{1:N}}
\newcommand{\datavarnd}{y_{1:N}(\delta)}
\newcommand{\datavarndone}{\datavarn(\delta_1)}
\newcommand{\datavarndtwo}{\datavarn(\delta_2)}
\newcommand{\datadone}{\mathcal{D}(\delta_1)}
\newcommand{\datadtwo}{\mathcal{D}(\delta_2)}

\section{Introduction}
In system identification and structural health monitoring, a common goal is to inform the parameters governing a predictive model of a structure using relevant data collected from sensors on it. For this purpose, it is often desirable to incorporate a Bayesian model updating framework to quantify uncertainty surrounding the model parameters based on sensor data. \citep{Beck1998, Beck2010, Vanik2000, Zhang2016}. For example, in a structural dynamics setting, the sensors would typically measure the physical response of the structure in terms of accelerations, strains, etc. under natural, induced or ambient excitation. The data is then used to inform the parameters of a model, which is typically a linear dynamical finite-element model with viscous damping but need not be. Any available prior information about the model parameters that is available can be incorporated in the Bayesian model updating framework, where \textit{prior} and \textit{posterior} probability distributions over parameters of interest are used to quantify uncertainty in their values \textit{before} and \textit{after} utilization of the data, respectively.

The optimal sensor placement problem may be viewed as optimally distributing sensors over a structure to maximize the information gained from the data about the model parameters, given a fixed sensor budget \citep{Beck1998a, Papadimitriou2000, Yuen2001, Papadimitriou2004, Yuen2015, Sun2015, Argyris2018}. For structures that are not simple geometric shapes, which is typically the case, the optimal sensor placement problem over all possible locations becomes difficult to specify. Therefore, only a pre-determined subset of locations is considered for instrumentation, for instance, the degrees of freedom of a finite-element model of the system. The optimization problem therefore becomes an inherently discrete configuration selection problem. In this case, heuristics such as genetic algorithms, greedy selection, swarm optimization etc., \citep{Papadimitriou2000} \citep{Papadimitriou2004} \citep{Sun2015} are usually needed in order to obtain a satisfactory sub-optimal solution, as the computational cost of a brute-force combinatorial search becomes prohibitive for realistic problems.

The optimality criterion is defined according to the purpose of collecting the sensor data. In this paper, it is assumed that the data would be used in a Bayesian model updating framework. In this setting, optimally placed sensors would collect data to maximize information gained about the model parameters. Then the mutual information \citep{Cover2005} between the model parameters and predicted data is a natural choice for the objective function for this optimization problem. Equivalently, the expected entropy of the posterior of the model parameters with respect to the predicted data can be chosen. This entropy-based objective function for optimal sensor placement was introduced in \citep{Beck1998a} \citep{Papadimitriou2000}.

Under the assumptions that imply applicability of the Laplace approximation to the posterior distribution, and assuming sufficiently small prediction errors, the entropy-based optimal sensor placement problem can be simplified to the optimization of the log-determinant of the Fisher information matrix (FIM) of the model parameters \citep{Beck1998a}. This gives us a rigorous information-theoretic foundation for one of the proposed choices of the objective function in \citep{Udwadia1994}, namely, the aforementioned log-determinant, rather than the trace, of the FIM. The main focus of this paper is to show how this log-determinant formulation can lend itself to a fast sub-optimal, and possibly optimal, solution using convex optimization \citep{Boyd2004} \citep{Joshi2009}, bypassing the need for a discrete optimization over all possible sensor configurations. The method is applied to an illustrative example of a 50-story uniform shear building, where there are over 47 trillion possible sensor configurations and the optimal sensor placement is found using fewer than 100 objective function evaluations.

\section{Bayesian model updating}
\label{sec:bayesmodelupdating}
This section lays out the framework for Bayesian model updating for structural dynamics which forms the basis for the optimal sensor placement problem. The measured data on which inference is to be performed is assumed to be available in the form of dynamic test data in the time domain. It is also assumed here that the complete system input can be determined given the parameters. In a typical structural dynamics setting, one observes the acceleration, velocity or displacement time history at a certain number of locations. The locations that are selected, could, for instance, correspond to a subset of the degrees-of-freedom of a finite-element model of the structure.

Denote the uncertain system model parameters by $\theta_s$. These could correspond, for instance, to sub-structure stiffnesses, Rayleigh damping parameters, etc. in the finite element model used to predict the response of the structure.

Denote by $y_n$ the stochastic prediction of the quantity to be observed at the time instant $t_n$. Note that $y_n \in \mathbb{R}^{N_o}$, where $N_o$ is the actual number of observed degrees-of-freedom out of a total of $N_d$ potential degrees-of-freedom for observation. The set of stochastic predictions for the observations over the whole duration of the measurements, $[t_1, t_N]$, is denoted by $\datavarn$.

The equation relating stochastic predictions of the observations, $y_n$, to the deterministic predictions from the model is:
\begin{equation}
\label{eqn:obspred}
y_n = S_o(\delta) \left( x_n(\theta_s) + \epsilon_n(\theta_e) \right)
\end{equation}
where the vector of deterministic predictions at each time-step of interest is denoted by $x_n(\theta_s) \in \mathbb{R}^{N_d}$; the uncertain \textit{prediction-error} is denoted by $\epsilon_n \in \mathbb{R}^{N_d}$; $\delta \in \{0,1\}^{N_d}$ is a binary vector that indicates the presence or absence of a sensor at a degree-of-freedom, so that the sum of entries of $\delta$ equals the number of observed DOFs, $N_o$; and the sensor selection matrix, $S_o(\delta) \in \mathbb{R}^{N_o \times N_d}$, selects the components that are to be observed from the full stochastic prediction. The uncertain prediction-errors are modeled by a probability distribution that depends on uncertain prediction-error parameters denoted by $\theta_e$; for example, using a zero-mean, discrete-time Gaussian white-noise process with a stationary and isotropic co-variance matrix having diagonal entries $\sigma^2$, $\theta_e$ would just be the scalar $\sigma$. For an anisotropic model for the co-variance matrix without prediction error correlation between different locations, $\theta_e$ would be a vector of variances corresponding to the diagonal entries of the co-variance matrix.

Denote the complete parameter vector by $\theta$, which is the collection of uncertain system parameters and prediction-error parameters, i.e., $\theta=[\theta_s, \theta_e]$. Prior information on the parameter values is expressed in terms of a \textit{prior distribution}, $p(\theta|\model)$. The \textit{stochastic forward model}, typically denoted by $p(\datavarn|\theta, \model)$, prescribes the probability of observing a particular data set given the parameters. The \textit{probability model class}, $\model$, is used to indicate a specific pairing of stochastic forward model and prior. The stochastic forward model with actual data $\datavarnhat$ substituted for $\datavarn$, when viewed simply as a function of the uncertain parameters, $\theta$, is called the \textit{likelihood function}. Our prediction-error modeling assumptions imply:
\begin{equation}
\label{eqn:likelihoodiid}
p(\datavarn|\theta, \model) = \prod_{n=1}^N p\left( y_n| \theta, \model \right)
\end{equation}

Bayes' rule implies that the posterior probability density, $p(\theta|\datavarn,\model)$, is proportional to the product of the likelihood function and prior probability density, thereby allowing an update of the prior:
\begin{equation}
\label{eqn:update}
p(\theta|\datavarn, \model)=\dfrac{p(\datavarn|\theta, \model)p(\theta|\model)}{p(\datavarn|\model)}
\end{equation}

\noindent Here, the normalizing factor $p(\datavarn|\model)$ is called the \textit{evidence} for the model class $\model$ provided by the data $\datavarn$. It can be determined using the Total Probability Theorem as:
%\noindent The normalizing factor here is called the \textit{evidence} for the model class $\model$ provided by the data $\datavarn$, is denoted by $p(\datavarn|\model)$ and which can be determined using the Total Probability Theorem as:
\begin{equation}
\label{eqn:totprobevd}
p(\datavarn|\model) = \int p(\datavarn|\theta, \model) p(\theta|\model) d\theta
\end{equation}

For a given data set, the parameter vector that maximizes the likelihood function is known as the maximum likelihood estimate (MLE), denoted by $\hat{\theta}_{MLE}$. Similarly, the parameter vector that maximizes the posterior probability density is known as the maximum \textit{a posteriori} (MAP) estimate of the parameters, denoted by $\hat{\theta}_{\text{MAP}}$. 

\subsection{The prediction-error model}
Based on the Principle of Maximum Information Entropy \citep{Beck2010} \citep{Cover2005}, we model the prediction errors as a discrete-time Gaussian process with equal variance at every degree of freedom, and uncorrelated across time and location, so that:
\begin{equation}
\label{eqn:gwn}
\mathbb{E}\left[\epsilon_n \epsilon_m^T \right] = \sigma^2 I_{N_d} \delta_{mn} \text{ where } \epsilon_n \sim \normal(\cdot|0, \sigma^2 I_{N_d}) \text{ i.i.d.}
\end{equation}
This implies that knowing the prediction-error at one time and location provides no information \textit{a priori} about the errors at other times and locations. This Gaussian distribution on the prediction errors gives the maximum entropy (most uncertainty) for a specified zero mean and common variance $\sigma^2$ for all times and locations. Combining Equations (\ref{eqn:obspred}), (\ref{eqn:likelihoodiid}) and (\ref{eqn:gwn}), the stochastic forward model then becomes:
\begin{equation}
\label{eqn:likelihoodiso}
p(\datavarn|\theta, \model) = \prod_{n=1}^N \normal \left(y_n|S_o(\delta) x_n(\theta), \sigma^2 \right)
\end{equation}

\noindent The logarithm of this is needed later:

\begin{equation}
\label{eqn:loglike}
\log p(\datavarn|\theta,\model) = -\dfrac{NN_o}{2} \log 2\pi \sigma^2 -\dfrac{1}{2\sigma^2} \sum_{n=1}^N \|y_n - S_o(\delta) x_n(\theta_s)\|^2
\end{equation}

\section{Mutual information-based optimal sensor placement}

The mutual information between two stochastic variables is a measure of about how much they imply about each other's possible values. It is a fundamental information-theoretic quantification of their correlation \citep{Cover2005}. In our case, the mutual information between the stochastic predictions for the observed data, $y_{1:N}$, and the parameter vector, $\theta$, is defined by:
\begin{flalign}
%\label{eqn:mi1}
I(\datavarnd, \theta|\model) &= \mathbb{E}_{\datavarnd,\theta|\model} \left[ \log \dfrac{p(\datavarnd, \theta|\model)}{p(\datavarnd|\model)p(\theta|\model)} \right] \label{eqn:mi1.1} \\
&= \int \int p(\datavarnd, \theta|\model) \left[ \log \dfrac{p(\theta|\datavarnd,\model)}{p(\theta|\model)} \right] d\datavarn d\theta \label{eqn:mi1.2} \\
&= \int p(\datavarnd|\model) \int p(\theta|\datavarnd, \model) \left[ \log \dfrac{p(\theta|\datavarnd,\model)}{p(\theta|\model)} \right] d\theta d\datavarn \label{eqn:mi1.3} \\
&= \mathbb{E}_{\datavarnd|\model} D_{KL}[p(\theta|\datavarnd,\model)||p(\theta|\model)] \label{eqn:mi1.4}
\end{flalign}
where $D_{KL}(\cdot||\cdot)$ is the Kullback-Leibler (KL) divergence (the inner integral over $\theta$ in Equaton (\ref{eqn:mi1.3})).

In the definition of mutual information in Equation (\ref{eqn:mi1.1}), the dependence of the stochastic prediction of the data $\datavarnd$ on the sensor configuration $\delta$ is made  explicit but will sometimes be omitted for convenience. Equations (\ref{eqn:mi1.2}) and (\ref{eqn:mi1.3}) come from the product rule in probability. These equations show that the mutual information can be expressed as the expected Kullback-Leibler (KL) divergence, or expected relative entropy, of the posterior with respect to the prior.

In the optimal sensor placement framework, we wish to maximize the mutual information between the predicted sensor data and the model parameters over all possible configurations of placing $N_o$ sensors at $N_d$ locations. This is equivalent to maximizing the expected KL divergence in Equation (\ref{eqn:mi1.4}), which implies that the optimal sensor configuration has the largest expected reduction in entropy of the posterior distribution from the prior. This corresponds to the greatest expected information gain about the parameters from the data during Bayesian updating. In other words, considering all sensor configurations, the optimal one would result in the maximum reduction in parameter uncertainty, on average, upon obtaining the data.

The mutual information in Equation (\ref{eqn:mi1.2}) can also be expressed in terms of two specified distributions specified by the model class $\model$, i.e., the prior and the stochastic forward model, by using the product rule again:

\begin{equation}
\label{eqn:miworkable}
I(\datavarnd; \theta|\model) = \int \int p(\datavarnd|\theta,\model) p(\theta|\model) \left[\log \dfrac{p(\datavarnd|\theta,\model)}{p(\datavarnd|\model)} \right] d\datavarnd d\theta
\end{equation}

\noindent The denominator in the logarithm term is what one would associate with the evidence for the model class $\model$, based on given data, as defined in Equation (\ref{eqn:totprobevd}). Here, the data is not available and this term is simply a function obtained upon marginalizing out the parameters from the joint on $y_{1:N}(\delta)$ and $\theta$.

%%%%%%%%%%%%%%%%%%%%%%% IMPORTANT MODIFICATION
%\section{Mutual information-based optimal sensor placement}
%\label{sec:ospmif}
Choosing the mutual information between the uncertain parameters and uncertain data as the objective function to be maximized over the sensor configurations, the optimal sensor placement problem becomes:

\begin{equation}
\label{eqn:generalprob}
\begin{aligned}
& \underset{\delta}{\text{maximize}}
& &I(\datavarnd; \theta) & \\
& \text{subject to}
& &  \delta_i \in \{0, 1\}, &\; i = 1, \ldots, N_d \\
& \text{and}
& & \mathds{1}^T \delta = N_o.&
\end{aligned}
\end{equation}

\noindent Equation (\ref{eqn:miworkable}) can be used for the objective function for the problem in (\ref{eqn:generalprob}). Recall that $\delta$ is a binary $N_d$-dimensional vector specifying the sensor configuration over the degrees of freedom in the system and so it has $N_o$ unit components and $N_d-N_o$ zero components. The explicit dependence on $\model$ has been suppressed in the notation in (\ref{eqn:generalprob}).

In principle, to perform the optimization in Equation (\ref{eqn:generalprob}), we need to search over all possible sensor configurations. Instead, we employ a convex optimization technique that greatly reduces the computational complexity of this search.

\section{Equivalent entropy-based optimal sensor placement}
\label{sec:entbasedopt}

In this section, an equivalent entropy-based optimal sensor placement problem is formulated and an efficient solution using convex optimization techniques is presented. The Laplace approximation to the posterior in a Bayesian system identification problem is described and used to express the entropy of the posterior.

Using the expression in Equation (\ref{eqn:mi1.2}) for mutual information:
\begin{equation}
\label{eqn:mutinfoentropy}
\begin{aligned}
I(y_{1:N}(\delta);\theta) &= \mathbb{E}_{y_{1:N}(\delta),\theta}\left[ \log \dfrac{p(\theta|y_{1:N}(\delta))}{p(\theta)} \right ] \\
&=H(p(\theta)) - \mathbb{E}_{y_{1:N}(\delta)} \left[ H(p(\theta|y_{1:N}(\delta))) \right]
\end{aligned}
\end{equation}
where $H(p(x)) = -\int p(x) \log p(x) dx$ is the entropy corresponding to $p(x)$. Here, the prior entropy term is irrelevant in the sensor placement problem and may be discarded to yield an objective function that depends only on the posterior entropy term:
\begin{equation}
\label{eqn:utilityfunction}
U(\delta) = \mathbb{E}_{y_{1:N}(\delta)}\left[ H(p(\theta|y_{1:N}(\delta))) \right ]
\end{equation}
%\log p(\theta|y_{1:N}(\delta))

\noindent Thus, the objective function maximization in Equation (\ref{eqn:generalprob}) can be replaced by an equivalent \textit{minimization} of the expected posterior entropy of the model parameters $\theta$. The idea of choosing the locations of the $N_o$ sensors so that the posterior uncertainty of the parameters is minimized was first introduced in \citep{Papadimitriou2000} \citep{Beck1998a}. In these papers, it is shown that this entropy-based criterion can be simplified to a more manageable log-determinant criterion when the following assumptions are made about the probability model class $\model$:
\begin{itemize}
\item The model class, upon collection of the data, is globally identifiable and the posterior distribution has a single peak \citep{Beck1998}.
\item The Laplace approximation holds; that is, a sufficient number of data points are measured so that the posterior is approximately Gaussian \citep{Beck1998}.
\end{itemize}

Applying the Gaussian approximation to the posterior, and using $\hat{\theta}(y_{1:N})$ to denote the MAP value of $\theta$ based on data $\datavarnd$:
\begin{equation}
\label{eqn:laplaceapprox}
p(\theta|y_{1:N}(\delta)) \approx \mathcal{N}(\theta|\hat{\theta}(y_{1:N}),A_N^{-1}(\hat{\theta}(y_{1:N}), \delta))
\end{equation}
where the precision matrix $A_N$ here is given by the negative of the Hessian of the logarithm of the stochastic forward model \citep{Beck1998}:
\begin{equation}
\label{eqn:hessiancomponents}
[A_N(\theta, \delta)]_{ij} = -\dfrac{\partial^2 \log p(y_{1:N}(\delta)|\theta)}{\partial \theta_i \theta_j}
\end{equation}
\noindent The objective function in Equation (\ref{eqn:utilityfunction}) is now simply related to the entropy of this approximately Gaussian posterior which may be expressed in terms of the log-determinant of its precision matrix as:
\begin{equation}
\label{eqn:mihessbasic}
U(\delta) = -\dfrac{1}{2}\mathbb{E}_{y_{1:N}(\delta)}\left[\log \det A_N(\hat{\theta}(y_{1:N}), \delta) \right]
\end{equation}

Note that the optimal sensor location problem is to be solved typically  \textit{before} any response data is available from the real structure to be instrumented, so predictions of the data must be made based on a probability model. This can be constrained based on available information about the structure,  either from its design or from preliminary tests, in the form of a structural model and a prior distribution over the system parameters, as shown in Section \ref{sec:bayesmodelupdating}. A nice feature of the log-determinant formulation above, is that the dependence on the data of the approximate posterior distribution over the uncertain parameters is only through the optimal parameters, $\hat{\theta} = \hat{\theta}(\datavarn)$, that is,
\begin{equation}
p(\theta_s|\datavarn, \model) = p(\theta_s|\hat{\theta}(\datavarn), \model)
\end{equation}

%\section{Log-determinant formulation}
%This section expands on the derivation of the log-determinant formulation set up the previous section. The Laplace approximation to the posterior together with additional assumptions about the prediction-errors, allows for the development of a log-determinant entropy-based objective.

The precision matrix $A_N$ in Equation (\ref{eqn:hessiancomponents}) of the multivariate Gaussian in Equation (\ref{eqn:laplaceapprox}) can be partitioned into two parts corresponding to the MAP values of the system parameters $\hat{\theta}_s \in \mathbb{R}^{N_m}$ and the prediction-error parameter, $\hat{\sigma} > 0$:
\begin{equation}
\label{eqn:parpart}
A_N(\theta_s) =  \begin{bmatrix} B_N(\theta_s) &0 \\ 
0 & C_N(\hat{\sigma}) 
\end{bmatrix}
\end{equation}

where

\begin{equation}
\label{eqn:bnmatrix}
[B_N(\theta_s)]_{ij} = -\begin{bmatrix} \dfrac{\partial^2 \log p(\datavarn|[\theta_s,\hat{\sigma}], \model)}{\partial \theta_{s,i} \partial \theta_{s,j}} \end{bmatrix}
\end{equation}
%\Bigg|_{\theta_s = \hat{\theta}_s}

and

\begin{equation}
\label{eqn:parpart3}
C_N(\hat{\sigma}) = -\dfrac{\partial^2 \log p(\datavarn|[\theta_s, \sigma], \model)}{\partial \sigma \partial \sigma} \Bigg|_{\sigma = \hat{\sigma}}
\end{equation}

\noindent The off-diagonal terms in the Hessian of Equation (\ref{eqn:parpart}) are zero because of our choice of stochastic forward model and prior. Also, in Equations (\ref{eqn:parpart}) - (\ref{eqn:parpart3}), $\hat{\sigma}^2$ for fixed system parameters $\theta_s$ is the MAP prediction-error variance, assuming a uniform prior on $\sigma^2$ over a large positive interval $(0, \sigma_{\max}^2)$:

\begin{equation}
\label{eqn:sigmahatsq}
\hat{\sigma}^2(\theta_s) = \dfrac{1}{NN_o} \sum_{n=1}^N \|y_n - S_o(\delta) x_n(\theta_s)\|^2 \stackrel{\Delta}{=} J(\theta_s)
\end{equation}

\noindent Equation (\ref{eqn:sigmahatsq}) can be used to re-express the stochastic forward model at the MAP variance parameter:
\begin{equation}
p(\datavarn|\theta=[\theta_s, \hat{\sigma}], \model) = [2\pi e J(\theta_s)]^{-NN_o/2}
\end{equation}

\noindent Also, the posterior distribution of the system parameters, with the prediction-error parameter marginalized out, can be expressed as:
\begin{equation}
\label{eqn:bmatrixintro}
p(\theta_s|\datavarn,\model) = \dfrac{\sqrt{\det B_N(\hat{\theta}_s)}}{(2\pi)^{NN_o/2}} \exp \left[ -\dfrac{1}{2} (\theta_s - \hat{\theta}_s)^T B_N(\hat{\theta}_s)(\theta_s - \hat{\theta}_s) \right]
\end{equation}

\noindent where from Equations (\ref{eqn:loglike}) and (\ref{eqn:bnmatrix}):

\begin{equation}
\label{eqn:bnbefore}
B_N(\hat{\theta}_s)=\dfrac{NN_o}{2J(\hat{\theta}_s)}\left[ \dfrac{\partial^2 J(\theta_s)}{\partial \theta_s \partial \theta_s} \right]_{\theta_s=\hat{\theta}_s}
\end{equation}

\noindent Furthermore, using Equation (\ref{eqn:parpart}), the log-determinant in Equation (\ref{eqn:mihessbasic}) can be expressed as:
\begin{equation}
\label{eqn:amatrixbc}
\log \det A(\hat{\theta}(y_{1:N}), \delta)=\log C_N(\hat{\sigma})+\log\det B_N(\hat{\theta}_s)
\end{equation}
where from Equations (\ref{eqn:loglike}) and (\ref{eqn:parpart3}):
\begin{equation}
C_N(\hat{\sigma}) = \dfrac{2NN_o}{\hat{\sigma}^2(\hat{\theta}_s)}
\end{equation}

\noindent Expanding the second-order derivative in Equation (\ref{eqn:bnbefore}) using the expression for $J(\theta_s)$ from Equation (\ref{eqn:sigmahatsq}):
\begin{equation}
\label{eqn:jexpanded}
[B_N(\hat{\theta}_s)]_{pq} = \dfrac{1}{\hat{\sigma}^2} \sum_{n=1}^N \left[ \dfrac{\partial x_n}{\partial \theta_{s,p}} S_o S_o^T \dfrac{\partial x_n^T}{\partial \theta_{s,q}}  +  \epsilon_n \left(S_o\dfrac{\partial^2 x_n}{\partial \theta_{s,p} \partial \theta_{s,q}}\right)^T  \right] \Bigg|_{\theta_s = \hat{\theta}_s}
\end{equation}

It is now assumed that the term involving the second derivative in Equation (\ref{eqn:jexpanded}) can be neglected. This can be justified if at least one of the following is true: the prediction errors, $\epsilon_n$, are small in magnitude; or the deterministic predictions, $x_n$, vary slowly with respect to the system parameters and therefore have small second derivatives. The curvature assumption can be assessed before instrumenting the structure, while the prediction-error assumption can only be validated after checking the agreement between the deterministic predictions and the data.

With this assumption, the relevant portion of the Hessian matrix can be expressed in terms of the prediction sensitivity coefficients at the observed degrees of freedom. The resulting approximation is then:
\begin{equation}
\label{eqn:approximate}
[B_N(\hat{\theta}_s)]_{pq} \approx \dfrac{1}{\hat{\sigma}^2} \sum_{n=1}^N \left[ \dfrac{\partial x_n}{\partial \theta_{s,p}} S_o S_o^T \dfrac{\partial x_n^T}{\partial \theta_{s,q}}    \right]
\end{equation}

\noindent In addition, notice the dependence on the sensor configuration becomes $S_o^TS_o = \text{diag}(\delta)$, so Equation (\ref{eqn:approximate}) can be transformed to a double-sum:

\begin{equation}
\label{eqn:approxsensor}
[B_N(\hat{\theta}_s)]_{pq} = \dfrac{1}{\hat{\sigma}^2} \sum_{i=1}^{N_d} \delta_i \sum_{n=1}^N \dfrac{\partial x_{n,i}}{\partial \theta_{s,p}} \dfrac{\partial x_{n,i}^T}{\partial \theta_{s,q}}
\end{equation}

\noindent Equation (\ref{eqn:approxsensor}) is a linear sum of contributions of terms from each sensor location. For convenience of notation in the formulation of the optimal sensor placement problem, we introduce the $N_m \times N_m$ matrix $Q=\hat{\sigma}^2 B_N(\hat{\theta}_s)$, so:
\begin{equation}
\label{eqn:Qfirsttime}
[B_N(\hat{\theta}_s)]_{pq} = \dfrac{1}{\hat{\sigma}^2}Q_{pq}(\delta; \hat{\theta}_s) = \dfrac{1}{\hat{\sigma}^2} \sum_{i=1}^{N_d} \delta_i Q_{pq}^{(i)} (\hat{\theta}_s)\end{equation}
where
\begin{equation}
\label{eqn:Qcomponents}
Q_{pq}^{(i)} = \sum_{n=1}^N \dfrac{\partial x_{n,i}}{\partial \theta_{s,p}} \dfrac{\partial x_{n,i}^T}{\partial \theta_{s,q}} \Bigg|_{\theta_s = \hat{\theta}_s}
\end{equation}

Having developed an approximate expression for the posterior distribution over the system parameters, we are now in a position to tackle the optimal sensor location problem.

To greatly simplify calculations, we use nominal values $\theta_{s0}$ and $\sigma_0^2$ for the system and variance parameters respectively, in place of the optimal parameters $\hat{\theta}_s$ and $\hat{\sigma}^2$. Then, from Equation (\ref{eqn:bmatrixintro}):
\begin{equation}
\label{eqn:priorhyperprior}
p(\theta_s|\theta_0, \model) = \dfrac{\sqrt{\det B_N(\theta_{s,0})}}{(2\pi)^{NN_o/2}} \exp \left[-\dfrac{1}{2}(\theta_s - \theta_{s,0})^T B_N(\theta_0) (\theta_s - \theta_{s,0}) \right]
\end{equation}

\noindent Since the designer is uncertain about the nominal values, a prior distribution, $p(\theta_0)$, where $\theta_0 = [\theta_{s,0}, \sigma_0^2]$, is specified for the nominal parameters. Therefore, Equation (\ref{eqn:priorhyperprior}) replaces the dependence of the posterior on unavailable data with prior information that is already available in terms of the nominal parameters. These parameters are either available in the form of structural design information, or in the form of information obtained from preliminary tests of the structure.

Thus, the objective function $U(\delta)$ in Equation (\ref{eqn:mihessbasic}) becomes, after using Equations (\ref{eqn:amatrixbc}), (\ref{eqn:Qfirsttime}) and dropping the constant terms and scale factor of $1/2$:
\begin{equation}
\begin{aligned}
h(\delta) &= -\mathbb{E}_{\theta_0} \left[ \log \det Q(\delta, \theta_0) \right] \label{eqn:utility2entropy} \\
&= -\int \log \det Q(\delta, \theta_{s,0}) p(\theta_{s,0}) d\theta_{s,0}
\end{aligned}
\end{equation}
where the expression for the matrix $Q(\delta, \theta_{s,0})$ is given in Equations (\ref{eqn:Qfirsttime}) and (\ref{eqn:Qcomponents}).

Thus, as in \citep{Papadimitriou2000}, the entropy-based optimal sensor location problem requires a minimization of the expected posterior entropy of the system parameters over the binary vector, $\delta$, to give the optimal sensor configuration, $\delta^\ast$:

\begin{equation}
\label{eqn:originalprob}
\begin{aligned}
& \underset{\delta}{\text{minimize}}
& &h(\delta) & \\
& \text{subject to}
& &  \delta_i \in \{0, 1\}, &\; i = 1, \ldots, N_d \\
& \text{and}
& & \mathds{1}^T \delta = N_o.&
\end{aligned}
\end{equation}

This is a combinatorial optimization problem, that is, we wish to minimize $h(\delta)$ by choosing $N_o$ sensor locations among $N_d$ possible locations. Depending upon the problem, this could become prohibitively expensive. Each sensor configuration requires one evaluation of the expectation of the log-determinant of a computationally expensive prediction. For example, selecting 20 locations from among 50 locations to instrument gives a number of possible combinations of about 47 trillion.

Clearly, a brute-force search for the optimal solution over all possible sensor configurations will often not be a feasible approach. However, it is usually the  case that it is not possible to guarantee an optimal solution without an exhaustive search. Heuristic methods such as genetic algorithms can produce an acceptable sub-optimal value if run for long enough. Incremental greedy algorithms are also guaranteed to produce a sub-optimal value within $(1-1/e)$ of the optimal \citep{Guestrin2005}, relative to the range from the least to the most optimal configuration. Here, we present an alternative approach along the lines of \citep{Joshi2009} that applies a convex relaxation technique to provide a sub-optimal solution to the problem.

\section{Convex relaxation of the combinatorial optimization}
\label{sec:convexrelax}

In this section, a relaxed version of the optimization problem is set up and argued to be convex. The relevant partial derivatives of the objective function, that is, its gradient and its Hessian matrix, are derived for the purpose of computing the solution. The final step involves replacing the expectation integral over the prior distribution on the uncertain parameters by its Monte Carlo approximation so that the problem may be solved by a generic convex solver. 

The original optimization problem in Equation (\ref{eqn:originalprob}) is relaxed to allow the components of the binary vector $\delta$ to take on any real values between zero and one. The problem thus specified turns out to be a convex optimization for which there are efficient algorithms to find the unique solution, as shown in the next subsection. However, we note that the solution to this new optimization problem need not be a binary vector, because it may contain entries between 0 and 1. The solution is still very meaningful, however. If the solution is indeed a binary vector, then it is \textit{the} optimal solution to the original combinatorial problem. If not, then it is an upper bound to the value of the objective function that also gives guidance to what the optimal solution is likely to be.

We replace the original binary sensor placement vector, $\delta$, by a vector, $z$, in the hypercube $[0, 1]^{N_d}$, resulting in the relaxed optimization problem:
\begin{equation}
\label{eqn:relaxed}
\begin{aligned}
& \underset{z}{\text{minimize}}
& & h(z) & \\
& \text{subject to}
& -&z_i \leq 0 &\\
& & & z_i \leq 1, &\; i = 1, \ldots, N_d \\
& \text{and}
& & \mathds{1}^T z = N_o.&
\end{aligned}
\end{equation}
where the objective function is a straightforward extension of the previous one in Equation (\ref{eqn:utility2entropy}):
\begin{equation}
\label{eqn:extobj}
h(z) = -\mathbb{E}_{\theta_s}\left[ \log \det Q(z, \theta_s) \right]
\end{equation}
by replacing $\delta_i$ with $z_i$ in the definition of $Q$ is Equation (\ref{eqn:Qfirsttime}).
A major advantage of the relaxed problem is that it allows one to use continuous, rather than discrete, optimization packages to solve the problem.

\subsection{Convex nature of the relaxed optimization}
Equation (\ref{eqn:relaxed}) describes a convex optimization problem where the objective function is convex in $z$, and the equality and inequality constraints are affine. The log-determinant function in Equation (\ref{eqn:extobj}) is convex in $z$ \citep{Joshi2009} and the expectation operator preserves the convexity of its argument. Thus, the problem has a unique global optimum. This can be determined computationally, as long as the objective function can be computed at every $z$.

For larger problems, it is useful to efficiently supply the gradient and Hessian of the objective function with respect to $z$. This avoids expensive computations of their finite-difference counterparts. Fortunately, their analytical expressions are tractable and, along with their Monte Carlo approximations, are described in Appendix \ref{sec:derobj}.

\subsection{Solver for the relaxed optimization}

Since the problem in Equation (\ref{eqn:relaxed}) is convex, it may be solved using Newton's method. In order to apply this in practice, numerical approximations to the objective and its derivatives may be used, as in Equations (\ref{eqn:objapprox}) through (\ref{eqn:hessapprox}). The MATLAB routine $\textit{fmincon}$ was used for the examples in this paper. Since the optimization problem has equality and inequality constraints, a constrained convex optimization solver needs to be employed. The expression for the Newton step is not as straightforward as in the unconstrained case and is given in \citep{Boyd2004}.

% *********************************************** EXAMPLE APPLICATION *****************************************************
\section{Application to multistory structures}
\label{sec:example}

\subsection{Problem description}
\label{sec:probdesc}

The problem of optimally placing a fixed number $N_o$ of sensors over a structure for system identification is considered. The structure is assumed to be well modeled by linear dynamics with classical viscous damping. For optimal design of the sensor layout, the structure is imagined to be subject to a sinusoidal ground acceleration at the base specified by an amplitude $a_0$ and frequency $\omega$. Therefore, the governing differential equation for its displacement vector $x(t)$ as a function of time is given by,
\begin{equation}
\label{eqn:MCK}
M \ddot{x}(t) + C \dot{x}(t) + K x(t) = - M \mathds{1} a_0 \sin(\omega t), \text{ with } x(0) = 0, \dot{x}(0) = 0
\end{equation}
where $M$, $C$ and $K$ are the $N_d \times N_d$ mass, damping and stiffness matrices respectively, and $\mathds{1}$ is a vector of 1's.

In this problem, a nominal natural frequency parameter, the Rayleigh damping parameters, input frequency and input amplitude are all taken to be uncertain, i.e. $\theta_s=[\omega_0, \alpha, \beta, \omega, a_0]$. The prediction-error variance $\sigma^2$ for a stationary, isotropic prediction error co-variance matrix, is the uncertain prediction-error parameter. The designer specifies a prior distribution on these uncertain parameters.

Before converting this system into its modal co-ordinates, $q_j(t)$, with corresponding natural frequency and damping ratio $\omega_j$ and $\zeta_j$ respectively, the problem is simplified to take into account uncertainties in modeling the system. This is done largely through a substructuring scheme. The assumptions result in a simple expression, relative to the uncertain parameters, for each mode.

The following simplifications are assumed:

\begin{enumerate}
\item $M = m M^\ast$ and $K = k K^\ast$, where the mass and stiffness matrices are factored into an uncertain scalar and a known matrix part. Scalars $k$ and $m$ will be called the uncertain nominal stiffness and mass of the system respectively.
\item $\omega_0 = \sqrt{\dfrac{k}{m}}$ is the uncertain nominal natural frequency parameter. Although $k$ and $m$ are uncertain, only their square-root ratio $\omega_0$ is relevant and its uncertainty may be treated explicitly.
\item $C = \alpha M + \beta K = \alpha m M^\ast + \beta k K^\ast$, is the Rayleigh damping matrix with uncertain parameters $\alpha$ and $\beta$. 
\end{enumerate}

As a consequence of these simplifying assumptions, the following conditions hold:

\begin{enumerate}
\item The modal frequencies $\omega_j = c_j \omega_0$, where $c_j^2$ are the known eigenvalues of the $(K^\ast, M^\ast)$ system.
\item The eigenvector matrix $\Phi$ that diagonalizes the $(K^\ast, M^\ast)$ system also diagonalizes the $(K, M)$ system and can be used to solve the original dynamical system in Equation (\ref{eqn:MCK}).
\item $\Phi^T M^\ast \Phi = \textup{diag}\{\mu_j \}$ and $\Phi^T K^\ast \Phi = \textup{diag} \{\kappa_j \}$, such that $\dfrac{\kappa_j}{\mu_j} = c_j^2$.
\end{enumerate}

Given these simplifications, a modal solution to the problem may be calculated analytically.

\subsection{Modal solution}
\label{sec:modesoln}
We let $x(t) = \Phi q(t)$, where vector $q(t)$ contains the modal displacement co-ordinates, so the original equation may be simplified to:

\begin{equation}
\Phi^T M \Phi \ddot{q}(t) + \Phi^T C \Phi \dot{q}(t) + \Phi^T K \Phi q(t) = \Phi^T M \mathds{1} a_0 \sin(\omega t)
\end{equation}

\noindent Under the assumptions in Section \ref{sec:probdesc}, this simplifies to a decoupled set of modal ordinary differential equations:

\begin{equation}
\ddot{q}_j(t) + 2\zeta_j\omega_j \dot{q}_j(t) + \omega_j^2 q_j(t) = a_j \sin(\omega t)
\end{equation}
where
\begin{align}
\omega_j^2 &= c_j^2 \omega_0^2 \\
2 \zeta_j \omega_j &= \alpha + \beta \omega_j^2 \\
a_j &= -\dfrac{a_0}{\mu_j} \Phi^T \mathds{1}
\end{align}

\noindent Each modal equation corresponds to a driven, damped oscillator whose solution is:

\begin{align}
\label{eqn:modesoln}
q_j(t) = a_j 
 & \left[ \left( \dfrac{\omega^3 + \omega_j^2 \omega (2\zeta_j^2 - 1)}{\omega_{d_j}} \right) \exp(-\zeta_j \omega_j t) \sin(\omega_{d_j} t) \right.  \\ 
&+ (2\zeta_j \omega_j \omega) \exp(-\zeta_j \omega_j t) \cos(\omega_{d_j} t)  + (\omega_j^2 - \omega^2) \sin(\omega t) \\
& \left. - (2\zeta_j \omega_j \omega) \cos(\omega t) \right]  /  \left[(\omega_j^2 - \omega^2)^2 + (2\zeta_j \omega_j \omega)^2 \right]
\end{align}

\noindent The prediction equation for the displacements then becomes,

\begin{equation}
x_i(t) = \mathlarger{\sum_{j = 1}^{N_d}} \Phi_{ij} q_j(t)
\end{equation}

\subsection{Sensor placement algorithm}
In order to solve the relaxed optimal sensor placement problem in this case, the following operations need to be performed:
\begin{enumerate}
\item Determine the eigenvalues $c_j^2$ and eigenvector matrix $\Phi$ of the $(K^\ast, M^\ast)$ system
\item Choose an initial guess $z_0$ for the sensor positions that satisfies the constraints
\item Generate $N_k$ samples of $\theta_s = [\omega_0, \alpha, \beta, a_0, \omega]$ from the designer-specified prior distribution $p(\theta_s)$ 
\item For each sample $\theta_s^{(k)}$, calculate the gradient of $q_j(t)$ with respect to $\theta_s$ for $t_n = n\Delta t \text{ for } n = 1, \ldots, N$ 
\item Calculate the gradient $\mathlarger{\nabla_{\theta_s}} x_i (t_n, \theta_s^{(k)})$
\item Calculate the elementary matrices $Q^{(i)}(\theta_s^{(k)})$ for every $i$ and $k$ and store them.
\item Compute the objective function, its gradient and Hessian at the current vector $z_m$ \label{step:repeat}
\item Update $z_m$ to $z_{m+1}$ and repeat steps \ref{step:repeat} and \ref{step:finish} until satisfaction of the convergence criterion \label{step:finish}
\end{enumerate}

\subsection{Illustrative example of uniform shear buildings}
\begin{figure}[!htbp]
\centering
\includegraphics[scale=5.0]{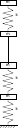}
\caption{Schematic of a uniform shear building}
\label{fig:springmassuniformshearbuilding}
\end{figure}

Consider the problem of placing a fixed number $N_o$ of displacement sensors on a $N_d$-DOF uniform shear building, as shown in Figure \ref{fig:springmassuniformshearbuilding}.

\FloatBarrier
In this case, the mass and stiffness matrices are given by:
\begin{align}
M &= mM^\ast = mI_{N_d} & \\
K &= kK^\ast = k 
\begin{bmatrix}
2 & -1 & 0 & \ldots & 0 & 0\\ 
-1 & 2 & -1 & 0 & \ldots & 0\\ 
0 & -1 & 2 & -1 & \ldots & 0 \\ 
 & \ddots & \ddots & \ddots & \ddots & \\ 
0 & \ldots & 0 & -1 & 2 & -1 \\
0 & 0 &\ldots & 0 & -1 & 1
\end{bmatrix} &
\end{align}

The prior parameters are taken to be independent and are distributed as:
\begin{flalign}
\omega_0 & \sim \ln \mathcal{N}(\cdot|\mu=2\pi,\sigma=0.25) & \\ 
\alpha & \sim \ln \mathcal{N}(\cdot|\mu=0.1,\sigma=0.01) & \\ 
\beta & \sim \ln \mathcal{N}(\cdot|\mu = 10^{-4},\sigma = 10^{-5}) & \\ 
a_0 & \sim \mathcal{N}(\cdot|\mu = 0, \sigma = 40\% \text{ gravitational acceleration}) & \\
\omega & \sim \ln \mathcal{N}(\cdot|\mu = 2\pi,\sigma = 0.25) & 
\end{flalign}

\noindent These choices are all reasonable and what is expected to be encountered in typical situations, although they have not been chosen to model any specific physical system. While the stochastic forward model is different for structures with a different number of stories, the same prior is used for each structure considered in this example.

\begin{table}[!htbp]
\centering
\caption{Simulation results}
\label{tab:sensor}
\begin{tabular}{|c|c|c|c|c|}
\hline
\multicolumn{4}{|c|}{\textbf{Case}}                                             & 
\multirow{2}{*}{\begin{tabular}[c]{@{}c@{}}\textbf{Sensor DOF \#}\\ (1 $\equiv$ Base, $N_d \equiv$ Roof)\end{tabular}} \\ \cline{1-4}
$N_d$ & $N_o$ & \multicolumn{1}{l|}{$N$}  & \multicolumn{1}{l|}{$N_k$} &                                                                                                               \\ \hline
2     & 1     & \multicolumn{1}{l|}{1000} & \multicolumn{1}{l|}{1000}  & 2                                                                                                             \\ \hline
2     & 1     & \multicolumn{1}{l|}{1000} & \multicolumn{1}{l|}{2000}  & 2                                                                                                             \\ \hline
2     & 1     & \multicolumn{1}{l|}{2000} & \multicolumn{1}{l|}{1000}  & 2                                                                                                             \\ \hline
2     & 1     & \multicolumn{1}{l|}{2000} & \multicolumn{1}{l|}{2000}  & 2                                                                                                             \\ \hline
4     & 1     & 1000                      & 1000                       & 4                                                                                                             \\ \hline
4     & 1     & 1000                      & 2000                       & 4                                                                                                             \\ \hline
4     & 1     & 2000                      & 1000                       & 4                                                                                                             \\ \hline
4     & 1     & 2000                      & 2000                       & 4                                                                                                             \\ \hline
4     & 2     & 1000                      & 1000                       & 2, 4                                                                                                          \\ \hline
4     & 2     & 1000                      & 2000                       & 2, 4                                                                                                          \\ \hline
4     & 2     & 2000                      & 1000                       & 2, 4                                                                                                          \\ \hline
4     & 2     & 2000                      & 2000                       & 2, 4                                                                                                          \\ \hline
8     & 2     & 1000                      & 1000                       & 6, 8                                                                                                          \\ \hline
8     & 2     & 2000                      & 2000                       & 6, 8                                                                                                          \\ \hline
\end{tabular}
\end{table}

Table \ref{tab:sensor} shows the results of solving the relaxed, convex optimization problem for structures with up to 8 degrees-of-freedom on which up to 2 sensors are to be installed. Once the $Q^{(i)}(\theta_s^{(k)})$ matrices have been computed and stored for samples $\theta_s^{(k)}$ of $\theta_s$, the optimization algorithm always converges to the same solution irrespective of the initial starting point, $z_0$. This is expected, since the problem is convex and is guaranteed to have a global minimum. These are exact optimal solutions since they give $z^\ast_i \in \{0, 1\}$ in each case. For each structure (value of $N_d$), the sensor configuration remains stable to changes in the number of time-steps $N$ or prior samples $N_k$ used (each chosen as either 1000 or 2000). There is a sensor on the roof in every case.

\begin{figure}[!htbp]
\centering
\includegraphics[scale=0.7]{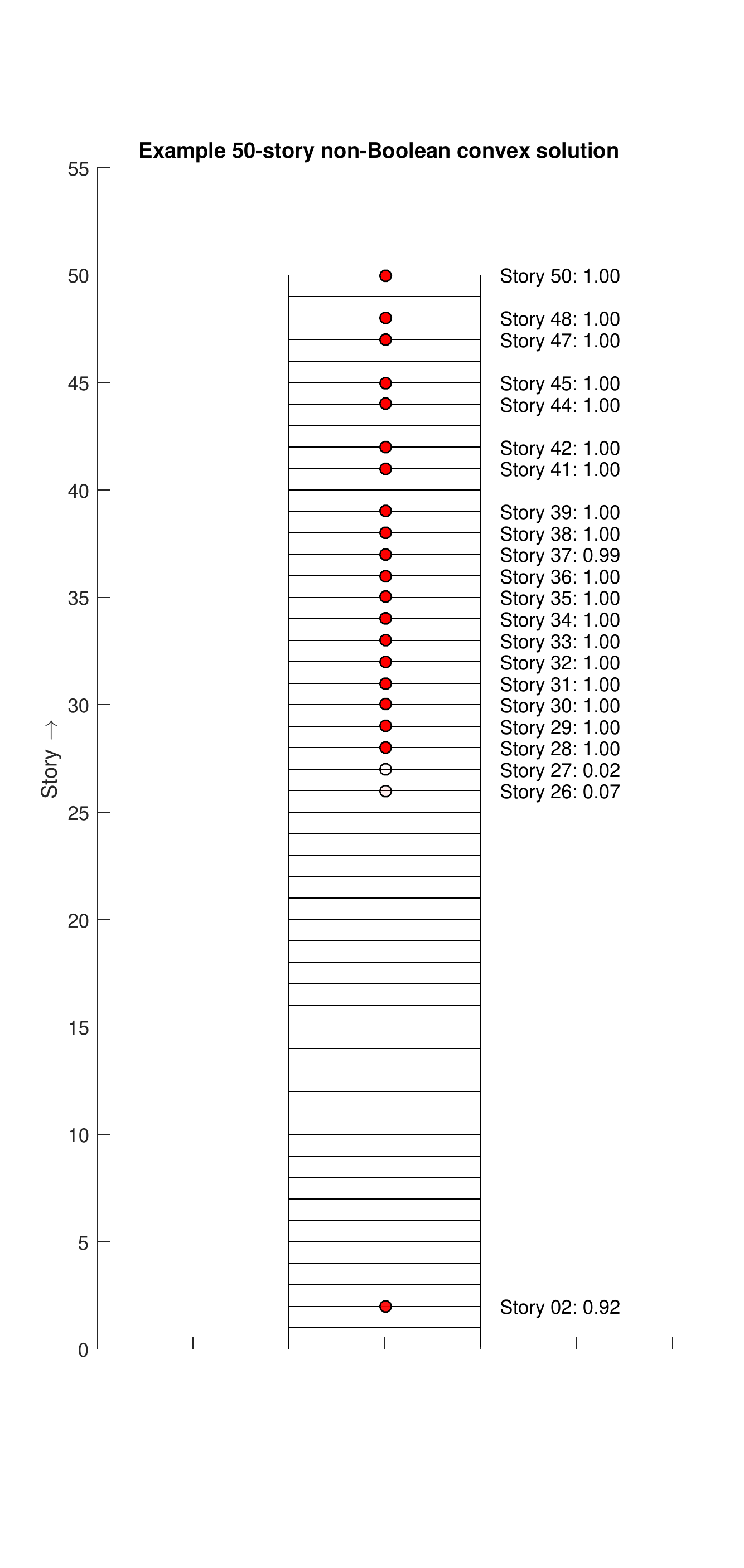}
\caption{The resulting sensor locations with their optimal value  $z^\ast_i$, after solving the convex optimization problem for a 50-story structure ($z_i^\ast=0$ at all locations without a sensor)}
\label{fig:50story1}
\end{figure}

\begin{figure}[!htbp]
\centering
\includegraphics[scale=0.7]{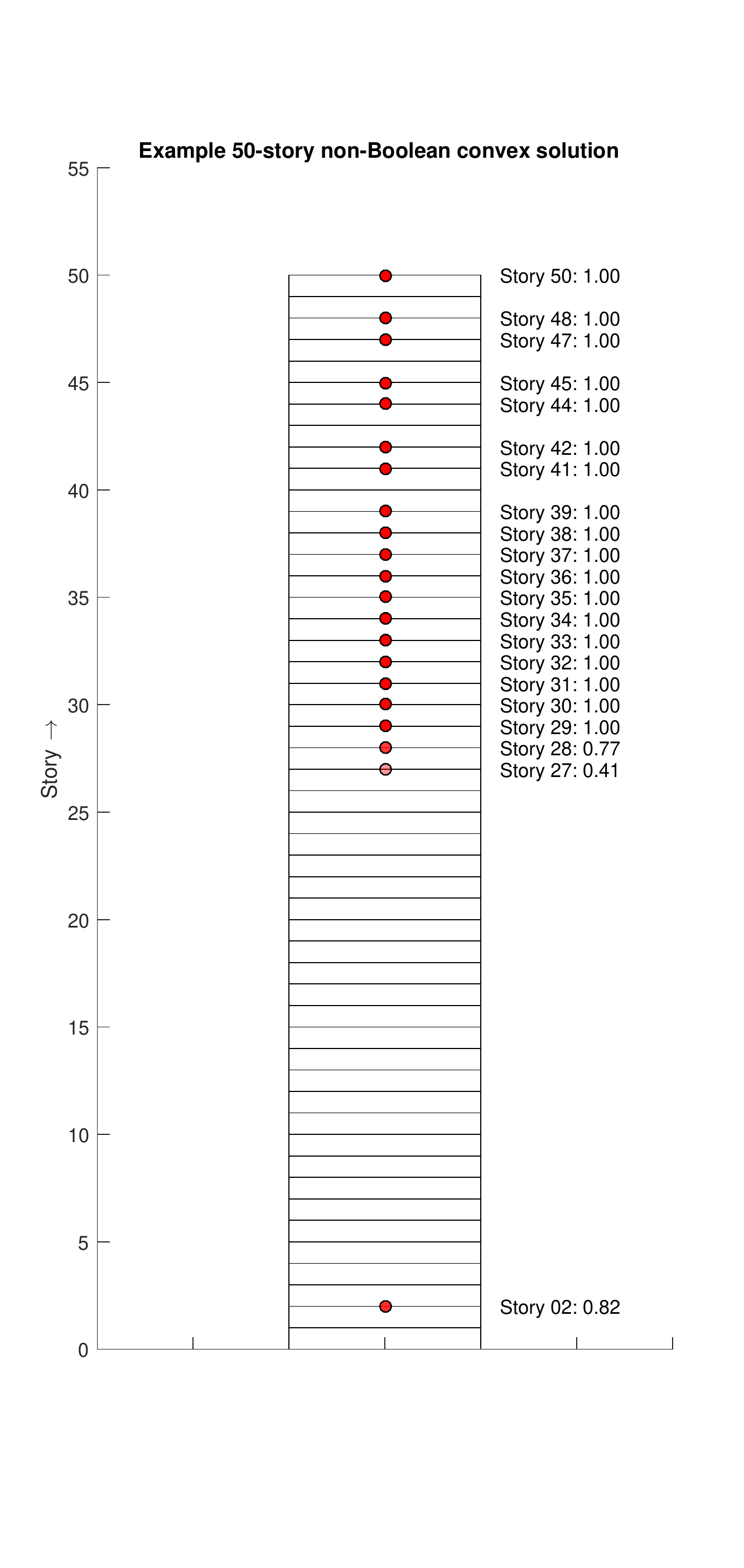}
\caption{Another solution for the 50-story building but with a different set of prior samples than in Figure \ref{fig:50story1}}
\label{fig:50story2}
\end{figure}

For a more challenging problem we consider the sensor location scheme for a 50-story shear building ($N_d=50$) to be instrumented with $N_o=20$ sensors. Some typical results are presented in Figures \ref{fig:50story1} and \ref{fig:50story2}. Since this problem involves a much larger number of degrees-of-freedom, there is a higher variance associated with the computation of the expectation of the log-determinant of the sensitivity matrix using prior samples. For this example, this shows up as a difference in the value of the non-binary sensor configuration at stories 26 through 28. It is to be noted, however, that there does not appear to be any ambiguity upon rounding the numbers to their binary values. For example, no sensors should be placed at stories 26 and 27 in Figure \ref{fig:50story1}, or at story 27 in Figure \ref{fig:50story2}.
\noindent The results show that all floors above the 27th, except floors 40, 43, 46 and 49, and no floor below the 28th except floor 2 should be instrumented.

\FloatBarrier

\FloatBarrier

\newpage 

As a reminder, an exhaustive search for the optimal configuration would have required over 47 trillion objective function evaluations. The gradient-based method requires fewer than 100 evaluations with a tolerance of $10^{-6}$ in MATLAB's \textit{fmincon} function. Of course, heuristic methods might have provided a good sub-optimal solution but in this example, optimal convergence is achieved by our method because the implied binary solution given by rounding the results in Figures \ref{fig:50story1} and \ref{fig:50story2} give essentially the same objective value as the optimum for the relaxed optimization problem.

\subsection{Comparison of differently chosen sensor configurations}

The number of bits of information gain of the optimal configuration, $z^\ast$, obtained using the convex technique over some other configurations is of interest here. Table \ref{tab:compsensdesc} describes the sensor configurations that are considered for the uniform 50-story structure:
\begin{table}[!htbp]
\centering
\caption{Description of various sensor configurations for comparative example}
\label{tab:compsensdesc}
\begin{tabular}{|c|c|}
\hline
\textbf{Case} & \textbf{Description} \\ \hline
$z^\ast$          & Optimal configuration using convex relaxation            \\ \hline
$z_{\text{low}}$         & Stories 1 through 20 instrumented            \\ \hline
$z_{\text{high}}$         & Stories 31 through 50 instrumented            \\ \hline
$z_{\text{common}}$          & Sensors evenly spaced (Stories 1 through 50 in steps of 2.5, rounded up)           \\ \hline
$z_{\text{greedy}}$         & Solution using greedy sequential placement            \\ \hline
\end{tabular}
\end{table}
\FloatBarrier
\noindent The configuration $z_{\text{greedy}}$ is the sensor configuration obtained using a greedy sequential placement algorithm that picks the best sensor location for the $(n+1)^\text{th}$ sensor after $n$ sensors are in place, starting with just one sensor. This algorithm is called the forward sequential sensor placement algorithm in \citep{Papadimitriou2004}.
The number of objective function evaluations is $N_o(2N_d - N_o) = 1600$ in this example, compared with fewer than 100 evaluations for the convex algorithm.

\begin{table}[!htbp]
\centering
\caption{Comparison of information gain relative to optimal configuration}
\label{tab:compsensinfo}
\begin{tabular}{|c|c|c|}
\hline
\textbf{Case} & \textbf{Objective Value} & \textbf{\# bits gain using z*} \\ \hline
$z^\ast$         & 6.14E+01                &                           \\ \hline
$z_{\text{low}}$         & 5.75E+01                & 5.7                            \\ \hline
$z_{\text{high}}$         & 6.12E+01                & 0.3                            \\ \hline
$z_{\text{common}}$         & 6.05E+01                & 1.3                            \\ \hline
$z_{\text{greedy}}$         & 6.14E+01                & 0.0                            \\ \hline
\end{tabular}
\end{table}
\FloatBarrier

\noindent Table \ref{tab:compsensinfo} shows that it is inefficient to place sensors in the lower stories relative to the higher ones.
\noindent It turns out in this case that $z_{\text{greedy}} = z^\ast$. With the convex relaxation scheme, however, it is immediately understood that the solution is optimal because it came out as a binary result (while the greedy solution is always binary but not necessarily optimal).

The commonly-used sensor configuration given by $z_{\text{common}}$ in Table \ref{tab:compsensdesc} appears to be significantly sub-optimal for the given model class. This configuration is sometimes popular, however, for considerations beyond the model class being considered; for instance, when it is expected that the behavior of the structure is not well understood, and so not well modeled.

\subsection{Variation of optimal configuration with parameter sample set}
For a Monte Carlo sample size, $N_k$, of 1000 or 2000, the optimal configuration is consistent for smaller structures ($N_d < 10$). For larger structures ($N_d \approx 50$), however, the optimal configuration may differ from sample set to sample set, presumably because the two sample sizes are not large enough for accurate estimation of the expectation. The severity of the problem increases with the size of the structure and the number of parameters used. It turns out that for this example, the majority of the sensor locations agree. Those in agreement are typically found in the upper half of the structure. This may change with the specific model choices made in formulating the problem, however.

Possible solutions to this problem:
\begin{itemize}
\item Select a sufficiently large sample size to avoid variation between sample sets
\item Avoid naive Monte Carlo integration and use an integral that provides an estimate with smaller variance. The method would depend on the nature of the integrand. An algorithm similar to simulated annealing could be used to compute the expectation value, such as in \citep{Beck2013}
\item Obtain better design information about the uncertain parameters to obtain a possibly more peaked prior with less variation between samples within a sample set, but this is not easy to accomplish
\end{itemize}

\subsection{Variation of optimal configuration with number of observation time-steps}
For a fixed parameter sample set, there is a variation in the optimal configuration when the number of observation time steps $N$ is changed from 1000 to 2000 for structures with $N_d > 10$. Sensor locations near the roof typically remain the same. However, below some story, the sensor locations can be quite different. This may be because of the significantly different amount of information provided about the dynamic response when the duration of measurement is doubled.

\section{Concluding remarks}
It was noted that the optimal sensor placement problem for system identification using any of the objective functions of (i) mutual information between the predicted sensor data and the system model parameters; (ii) the expected information gain about the parameters from the data; or (iii) the expected posterior uncertainty (entropy) of the parameters, give equivalent results.

Then the original combinatorial entropy-based minimization problem for optimal sensor placement was cast into a relaxed convex optimization problem and solved efficiently to yield sub-optimal, and sometimes optimal, solutions. Locations for which a sensor was unambiguously present or absent were typically obtained, across differing Monte Carlo samples for performing a required expectation and differing number of sample time steps for the data. Some partly ambiguous locations occurred where sample variability led to slightly different results. Ambiguous locations could be resolved using a brute-force combinatorial approach since the problem size for optimizing over just the ambiguous locations would be reduced and manageable.

In future work, it would be beneficial to explore the effects on the optimal sensor locations of spatial correlation between the predicted data at the possible sensor locations. Such a study could build on the preliminary investigation in Chapter IV of \citep{SelfThesis}.

%%%%%%%%%%%%%%%%%%%%%%%%%%%%%%%%%%%%%%%%%%
\vspace{6pt} 

\appendixtitles{no} %Leave argument "no" if all appendix headings stay EMPTY (then no dot is printed after "Appendix A"). If the appendix sections contain a heading then change the argument to "yes".
\appendixsections{multiple} %Leave argument "multiple" if there are multiple sections. Then a counter is printed ("Appendix A"). If there is only one appendix section then change the argument to "one" and no counter is printed ("Appendix").
\appendix
\section{Derivatives of the objective function}
%\unskip
%\subsection{}
\label{sec:derobj}

\subsection{Gradient of the objective}
%Identities in Appendix \ref{apx:matrix} are used. 

\begin{align}
\dfrac{\partial h}{\partial z_i} (z) &= -\dfrac{\partial}{\partial z_i}\mathlarger{\int} \log \det Q(z, \theta_s) p(\theta_s) d\theta_s & \\
&= -\mathlarger{\int} \dfrac{\partial \log \det Q}{\partial z_i}  p(\theta_s) d\theta_s & \text{(interchange)} \\
&= -\mathlarger{\int} \text{tr}\left( Q^{-1} \dfrac{\partial Q}{\partial z_i} \right) p(\theta_s) d\theta_s & \text{(derivative of log-det)} \\
&= -\mathlarger{\int} \text{tr}\left( Q^{-1}(z,\theta_s) Q^{(i)}(\theta_s) \right) p(\theta_s) d\theta_s & \text{(evaluation of partial)}
\end{align}
\begin{align}
\;\;\;\;\;\;\;\;\;&= -\mathlarger{\int} \left( \mathlarger{\sum_{j=1}^{N_a} \sum_{k=1}^{N_a}} Q^{-1}_{jk}(z,\theta_s) Q_{kj}^{(i)}(\theta_s) \right) p(\theta_s) d\theta_s  & \text{(trace of product)} \\
&= -\mathlarger{\int} \left( \mathlarger{\sum_{j=1}^{N_a} \sum_{k=1}^{N_a}} Q^{-1}_{jk}(z,\theta_s) Q_{jk}^{(i)}(\theta_s) \right) p(\theta_s) d\theta_s  & \text{(symmetry)} \\
&= -\mathbb{E}_{\theta_s} \left[ Q^{-1}(z,\theta_s)\mathbf{:}Q^{(i)}(\theta_s) \right]  & \text{(notation)} \label{eqn:gradientQz}
\end{align}

\subsection{Hessian of the objective}

\begin{align}
\dfrac{\partial^2 h}{\partial z_p \partial z_q}(z) &= -\dfrac{\partial}{\partial z_q} \mathlarger{\mathbb{E}_{\theta_s}} \left[ \mathlarger{\sum_{j=1}^{N_a} \sum_{k=1}^{N_a}} Q^{-1}_{jk}(z,\theta_s) Q_{kj}^{(p)}(\theta_s) \right] & \\
&= - \mathlarger{\mathbb{E}_{\theta_s}} \left[ \mathlarger{\sum_{j=1}^{N_a} \sum_{k=1}^{N_a}} \dfrac{\partial Q^{-1}_{jk}(z,\theta_s)}{\partial z_q} Q_{kj}^{(p)}(\theta_s) \right] & \text{(interchanges)}\\
&= \mathlarger{\mathbb{E}_{\theta_s}} \left[  \mathlarger{\sum_{j=1}^{N_a} \sum_{k=1}^{N_a}} \left( Q^{-1} \dfrac{\partial Q}{\partial z_q} Q^{-1} \right)_{jk}(z,\theta_s) Q_{jk}^{(p)}(\theta_s) \right] & \text{(derivative of inverse)} \\
&= \mathlarger{\mathbb{E}_{\theta_s}} \left[  \text{tr}  \left( Q^{-1}(z) Q^{(q)} Q^{-1}(z) Q{(p)} \right) (\theta_s) \right]  & \text{(derivative, trace)} \\
&= \mathlarger{\mathbb{E}_{\theta_s}} \left[  ((Q^{-1}(z) Q{(q)})\mathbf{:}((Q^{-1}(z) Q{(p)})^T) (\theta_s) \right]  & \text{(associativity, trace)} \label{eqn:hessianQz}
\end{align}

\subsection{Numerical approximations}
\label{sec:derobjapprox}
Given $N_k$ samples $\theta_s^{(k)}$ distributed according to the prior $p(\theta_s)$ specified by the designer of the sensor installation, the expectation integrals in Equations (\ref{eqn:extobj}), (\ref{eqn:gradientQz}) and (\ref{eqn:hessianQz}) may be approximated by their corresponding Monte Carlo estimates.

For the objective function,
\begin{equation}
\label{eqn:objapprox}
h(z) \approx -\dfrac{1}{N_k} \mathlarger{\sum_{k=1}^{N_k}} \log \det Q(z, \theta_s^{(k)})
\end{equation}

For the gradient of the objective function,
\begin{equation}
\label{eqn:gradapprox}
\dfrac{\partial h}{\partial z_i} (z) \approx -\dfrac{1}{N_k} \mathlarger{\sum_{k=1}^{N_k}} \mathlarger{\sum_{j=1}^{N_a} \sum_{k=1}^{N_a}} Q^{-1}_{jk}(z,\theta_s^{(k)}) Q_{jk}^{(i)}(\theta_s^{(k)}) 
\end{equation}

Finally, for the Hessian of the objective function,

\begin{equation}
\label{eqn:hessapprox}
\dfrac{\partial^2 h}{\partial z_p \partial z_q}(z) \approx \dfrac{1}{N_k} \mathlarger{\sum_{k=1}^{N_k}}   \mathlarger{\sum_{j=1}^{N_a} \sum_{k=1}^{N_a}}  \left( Q^{-1} Q{(q)} Q^{-1} \right)_{jk}(z, \theta_s^{(k)}) Q{(p)}(\theta_s^{(k)})_{kj} 
\end{equation}

Some computational effort may be spared by noting that,
\begin{align}
Q(z, \theta_s) &= \mathlarger{\sum_{l=1}^{N_d}} z_l \mathlarger{\sum_{n=1}^{N}} \dfrac{\partial x_l}{\partial \theta_s} \dfrac{\partial x_l}{\partial \theta_s}^T (t_n, \theta_s) \\
&= \mathlarger{\sum_{l=1}^{N_d}} z_l Q^{(l)}(\theta_s)
\end{align}

Hence, stored values of $Q^{(i)}(\theta_s)$ may be used to determine $Q(z, \theta_s)$.

%%%%%%%%%%%%%%%%%%%%%%%%%%%%%%%%%%%%%%%%%%
\vspace{6pt}

%%%%%%%%%%%%%%%%%%%%%%%%%%%%%%%%%%%%%%%%%%

%=====================================
% References, variant B: external bibliography
%=====================================
\externalbibliography{yes}
\bibliography{Mendeley}

%%%%%%%%%%%%%%%%%%%%%%%%%%%%%%%%%%%%%%%%%%
%% optional
%\sampleavailability{Samples of the compounds ...... are available from the authors.}

%% for journal Sci
%\reviewreports{\\
%Reviewer 1 comments and authors’ response\\
%Reviewer 2 comments and authors’ response\\
%Reviewer 3 comments and authors’ response
%}

%%%%%%%%%%%%%%%%%%%%%%%%%%%%%%%%%%%%%%%%%%
\end{document}